\documentclass[preprint,aps]{revtex4}
\usepackage{graphicx}
\usepackage{dcolumn}
\usepackage{bm}
\usepackage{amssymb}
\usepackage{amsmath}
\begin{document}
\setcounter{page}{1}
\title 
{A non commutative model for a mini black hole}
\author
{I. Arraut$^1$, D. Batic$^2$ and M. Nowakowski$^1$}
\affiliation{$^1$  
Departamento de Fisica, Universidad de los Andes, 
Cra.1E No.18A-10, Bogota, Colombia\\
$^2$ Departamento de Matematica, Universidad de los Andes,
Cra 1E, No. 18A-10, Bogota, Colombia}







\begin{abstract}
We analyze the static and spherically symmetric perfect fluid solutions 
of Einstein field equations inspired by the non commutative geometry. In the framework
of the non commutative geometry this solution is interpreted as a mini black hole
which has the Schwarzschild geometry outside the event horizon, but whose standard central singularity
is replaced by a self-gravitating droplet. 
The energy-momentum tensor of the
droplet is of the anisotropic fluid obeying a nonlocal equation of state.
The radius of the droplet is finite and the
pressure, which gives rise to the hydrostatic equilibrium, is  positive definite in the interior.
\end{abstract}


\maketitle

\section{Introduction}
In the last decades much effort has been invested to understand the
quantum effects in and of black holes. Beside the standard area of studying
the Hawking radiation under different situations \cite{page}, the question
regarding the final fate of a black hole, related to the problem
of the central singularity and a possible black hole remnant, has attracted much attention \cite{adler,horo,koch,ahn, we}.
One of the candidates of the mathematical framework for quantum gravity is the non-commutative geometry
\cite{Nicolini2,Spal1,Spal2,Piero,Rizzo,Ansoldi,Casadio,Spallucci} based on the non-commutativity of the
coordinates, $[x^{\mu}, x^{\nu}]=\theta^{\mu \nu}$, which is assumed to
be important at Planckian scales. Due to the uncertainty relation
$\Delta x^{\mu}\Delta x^{\nu} \ge 1/2 |\theta^{\mu \nu}|$,
the expectation value of the coordinates becomes smeared which effectively  can be
interpreted as a mass distribution $\rho_{\theta}$.
In \cite{Ansoldi}  an explicit model of a micro black hole has been constructed, based upon the fact that
$\rho_{\theta}$ is given by a Gaussian distribution.. In addition to that an equation
of state (EOS) for the radial pressure was assumed to be of the form, $p_r=-\rho_{\theta}$.
Insisting on a hydrostatical equilibrium, expressed through the Tolman-Oppenheimer-Volkov (TOV) equation,
and an energy-momentum tensor of a perfect fluid such a system would be clearly
over-determined as the density is already given by $\rho_{\theta}$ and the pressure by
the EOS. The remedy offered in \cite{Ansoldi} is to assume an anisotropic fluid with an additional tangential 
pressure.
The functional form of the latter is determined by the TOV equation. In this way, a
self-gravitating droplet is constructed, however at the price that the radial and tangential
pressures are always negative, a fact which is difficult to interpret.
In the present paper we therefore set out to overcome this difficulty
and attempt to construct a non commutative mini black hole model in which
the pressure is positive definite in the interior of the droplet. To this end, we have to choose
an appropriate EOS and the energy-momentum tensor. One could also proceed without the EOS
and use the TOV equation to determine the radial pressure. However, it is known that e.g. 
assuming the density to be constant, in such a case the radius $R$ of the extension of the
self-gravitating body  comes out to be $R \ge(9/8)r_s$ where $r_s$ is the Schwarzschild radius.
This illustrates the difficulty to construct a self-gravitating model with $R$ less or equal the
Schwarzschild radius. Therefore we adopt a non-local EOS used in different context in relativistic
astrophysics  \cite{Percoco,Hern,Hern2}. We then show that using a perfect fluid  energy-momentum tensor
and a special ansatz for the metric, the TOV equation is identically satisfied. This solves the problem
that the system might be over-determined. However, it can be shown that the interior metric cannot
be matched to the exterior Schwarzschild geometry. This forces us to modify the perfect fluid
energy-momentum tensor to  allowing some anisotropy. The resulting energy-momentum tensor,
which besides a radial pressure contains also a tangential one,  is well-known
and used also in various situations  \cite{Nicolini2,Stewart,Bayin,Madsen,Letelier,Haggag}. The whole model then proves successful in the following sense:
\begin{itemize}
\item[(i)] A self-gravitating droplet exists with a finite radius equal the Schwarzschild radius.
\item[(ii)] The interior metric can be matched to the Schwarzschild geometry outside.
\item[(iii)] The radial and tangential pressure are positive definite inside the object.
\end{itemize}


\section{Solution with a nonlocal equation of state}
According to \cite{Spal1,Spal2} non commutativity turns point-like structures into smeared objects: this is achieved by replacing the position Dirac delta with a Gauss distribution of standard deviation $\sigma=\sqrt{2\theta}$. In particular, we consider a particle-like gravitational source whose  mass density is static, spherically symmetric and it is given by \cite{Ansoldi}
\begin{equation}\label{density}
\rho_{\theta}(r)=\frac{M}{(4\pi\theta)^{3/2}}~e^{-r^2/(4\theta)}
\end{equation}
where $M$ denotes the total mass and $\theta$ is a positive parameter encoding non commutativity. However, as pointed out in \cite{Nicolini2} non commutativity plays a role on a scale $\sqrt{\theta}<10^{-16}$ cm. This implies that the mass distribution of our object is described by a Gaussian with a very narrow peak. In the following we investigate static, spherically symmetric solutions of Einstein equations where (\ref{density}) describes the energy density of the system. Since the parameter $\theta$ is so tiny such solutions are to be considered as a sort of microscopic solutions.  
In particular, we model the source term by means of a spherically symmetric perfect fluid which by definition satisfies the pressure isotropy condition, that is the radial and angular directional pressures coincide. Moreover, we suppose that our fluid obeys the nonlocal equation of state \cite{Percoco,Hern,Hern2}
\begin{equation}\label{NLES}
P(r)=\rho_\theta(r)-\frac{2\sqrt{\theta}}{r^3}\int_0^r u^2\rho_\theta(u)~du
\end{equation}
where $P$ is the pressure and that the energy-momentum tensor is assigned through
\begin{equation}\label{EMT}
T^{\mu}{}_{\nu}={\rm{diag}}(\rho_\theta,-P,-P,-P).
\end{equation}
As it is evident from (\ref{EMT}), we are assuming that the spatial velocities $u^i$ of the fluid with $i=1,2,3$ vanish since we are interested in matter configurations at hydrostatic equilibrium. 
We can determine the extension $R$ of the object by
the condition $P(R)=0$. The numerical value of $R$ can be determined by plotting  $\theta P$ vs. $r^{*}=r/\sqrt{\theta}$. 
To this purpose, it is convenient to introduce the new radial variable $r^{*}$ and the mass variable $\mu=M/\sqrt{\theta}$ so that (\ref{NLES}) becomes
\[
\theta P(r^{*})=\frac{\mu}{\pi^{3/2}}\left(\frac{e^{-{r^{*}}^2/4}}{8}-\frac{1}{{r^{*}}^3}\int_0^{{r^{*}}^2/4}s^{1/2}e^{-s}ds\right).
\]
The value at which the positive definite 
pressure vanishes is calculated numerically to be $R^{*}=R/\sqrt{\theta}\approx 1.269$. Let us consider a line element of the form
\[
g_{\mu\nu}={\rm{diag}}\left(e^{2\nu},-e^{2\lambda},-r^2,-r^2\sin^2{\vartheta}\right) 
\]
with $\nu=\nu(r)$ and $\lambda=\lambda(r)$. The energy-momentum tensor $T^{\mu\nu}$ satisfies the conservation condition
\[
0=T^{\mu\nu}{}_{;\nu}=\partial_{\nu}T^{\mu\nu}+\Gamma^{\mu}_{\nu\lambda}T^{\lambda\nu}+\Gamma^{\nu}_{\nu\lambda}T^{\mu\lambda}.
\]
If we take $\mu=r$ in the above equation a boring but straightforward computation gives
\[
\partial_r T^{r}{}_{r}=-\frac{1}{2}g^{00}\partial_r g_{00}\left(T^{r}{}_{r}-T^{0}{}_{0}\right)-\frac{1}{2}g^{\vartheta\vartheta}\partial_r g_{\vartheta\vartheta}\left(T^{r}{}_{r}-T^{\vartheta}{}_{\vartheta}\right)+
\]
\begin{equation}\label{emt}
-\frac{1}{2}g^{\varphi\varphi}\partial_r g_{\varphi\varphi}\left(T^{r}{}_{r}-T^{\varphi}{}_{\varphi}\right).
\end{equation}
In view of (\ref{EMT}) equation (\ref{emt}) gives rise to the following first order linear ODE for the pressure, namely
\begin{equation}\label{ODE_P}
P^{'}+\nu^{'}\left(\rho_\theta+P\right)=0.
\end{equation}
From the Einstein field equations we obtain the following three equations (the fourth equation coincides with the third one up to a multiplicative factor)
\begin{eqnarray}
&&\frac{1}{r^2}+\frac{e^{-2\lambda}}{r}\left(2\lambda^{'}-\frac{1}{r}\right)=8\pi\rho_\theta,\label{uno}\\
&&\frac{1}{r^2}-\frac{e^{-2\lambda}}{r}\left(2\nu^{'}+\frac{1}{r}\right)=-8\pi P,\label{due}\\
&&e^{-2\lambda}\left(\frac{\lambda^{'}}{r}-\frac{\nu^{'}}{r}-\nu^{''}+\nu^{'}\lambda^{'}-{\nu^{'}}^2\right) =-8\pi P.\label{tre}
\end{eqnarray}
From (\ref{uno}) we can compute the component $g_{rr}$ of the metric. In fact, (\ref{uno}) can be rewritten as
\begin{equation}\label{quattro}
\frac{d}{dr}\left(\frac{r}{e^{2\lambda}}\right)=1-8\pi\rho_\theta r^2.
\end{equation}
Integrating (\ref{quattro}) we obtain
\begin{equation}\label{cinque}
\frac{r}{e^{2\lambda}}=r-2M(r)+B
\end{equation}
with $B$ integration constant and
\begin{equation}\label{gamma}
M(r)=4\pi\int_0^r u^2\rho_\theta(u)du=\frac{2M}{\sqrt{\pi}}~\gamma\left(\frac{3}{2},\frac{r^2}{4\theta}\right)
\end{equation}
where $\gamma$ denotes the lower incomplete gamma function. Finally, from (\ref{cinque}) we have
\[
e^{2\lambda}=\frac{1}{1-\frac{2M(r)}{r}+\frac{B}{r}}.
\]
The main difficulty is the computation of the component $g_{00}$ of the metric. Following \cite{Hern} we introduce the new variables
\[
e^{2\nu}=h(r)e^{4\beta(r)},\qquad e^{2\lambda}=\frac{1}{h(r)}.
\]
As a consequence (\ref{uno}), (\ref{due}) and (\ref{tre}) become
\begin{eqnarray}
&&\frac{1-h-h^{'}r}{r^2}=8\pi\rho_\theta,\label{H1}\\
&&-\frac{1-h-h^{'}r}{r^2}+\frac{4h\beta^{'}}{r}=8\pi P,\label{H2}\\
&&\frac{h^{'}+2h\beta^{'}}{r}+\frac{1}{2}\left(h^{''}+4h\beta^{''}+6h^{'}\beta^{'}+8h{\beta^{'}}^2\right) =8\pi P.\label{H3}
\end{eqnarray}
By rewriting (\ref{NLES}) as an ODE, namely
\begin{equation}\label{L1}
\rho_\theta-3P+r(\rho_\theta^{'}-P^{'})=0
\end{equation}
and employing (\ref{H1}) and (\ref{H2}) we obtain the following second order linear ODE for $\beta$
\[
\frac{2}{r}\left(h^{'}+2h\beta^{'}\right)+h^{''}+2h^{'}\beta^{'}+2h\beta^{''}=0
\]
whose general solution is
\begin{equation}\label{M1}
\beta(r)=-\frac{1}{2}\ln{h}+C_1\int\frac{dr}{r^2 h}+C_2.
\end{equation}
One of the integration constants can be fixed by requiring that the above solution is consistent with the nonlocal equation of state we are working with. In fact, if we rewrite (\ref{H2}) by means of (\ref{H1}) and we make use of (\ref{M1}) we obtain
\[
P(r)=-\rho_\theta+\frac{h\beta^{'}}{2\pi r}=\rho_\theta-\frac{2\sqrt{\theta}}{r^3}\int_0^r u^2\rho_\theta(u)+\frac{C_1}{4\pi r^3}.
\]
Thus, consistency with (\ref{NLES}) requires that $C_1=0$. Therefore, we conclude that $e^{2\nu}=A/h$ with a positive constant $A=e^{4C_2}$ and we end up with the new line element
\begin{equation}\label{line_element}
ds^2=\frac{1}{h(r)}\left(A~dt^2-dr^2\right)-r^2d\vartheta^2-r^2\sin^2{\vartheta}~d\varphi^2
\end{equation}
describing the manifold $\mathbb{R}\times[0,R]\times S^2$. At a first sight it is not clear if the chosen NLES is compatible with the Tolman-Oppenheimer-Volkov (TOV) equation for the problem under consideration. Taking into account that the corresponding TOV equation in the present case is
\begin{equation}\label{TOV}
P^{'}=-(\rho_\theta+P)\frac{M(r)+4\pi r^3P}{r(r-2M(r))}.
\end{equation}
it is not difficult to verify that the nonlocal equation of state (\ref{NLES}) is consistent with (\ref{TOV}). To this purpose, it is convenient to rewrite (\ref{H1}), (\ref{H2}) and (\ref{H3}) in terms of the mass function $M(r)$ as follows \cite{Hern2} 
\begin{eqnarray}
&&\frac{M^{'}}{r^2}=4\pi\rho_\theta,\label{T1}\\
&&\frac{M^{'}}{r^2}-\frac{2M}{r^3}=4\pi P,\label{T2}\\
&&\frac{M^{''}}{r}+\frac{2(M^{'}r-M)}{r^3}\left(\frac{M^{'}r-M}{r-2M}-1\right)=8\pi P.\label{T3}
\end{eqnarray}
Taking into account that in terms of $M(r)$ the equation (\ref{NLES}) becomes
\begin{equation}\label{pressure}
P=\rho_\theta-\frac{M}{2\pi r^3}
\end{equation}
we find that 
\begin{equation}\label{A}
P^{'}=\frac{M^{''}}{4\pi r^2}+\frac{3M}{2\pi r^4}-\frac{M^{'}}{\pi r^3}
\end{equation}
where we used (\ref{T1}). On the other hand by means of (\ref{T1}) and (\ref{T2}) the r.h.s. of the TOV equation can be rewritten as
\begin{equation}\label{B}
-(\rho_\theta+P)\frac{M(r)+4\pi r^2P}{r(r-2M(r))}=-\frac{(M^{'}r-M)^2}{2\pi r^4(r-2M)}.
\end{equation}
The last step is to bring the r.h.s. side of (\ref{A}) to the form (\ref{B}). What we need is to express the second order derivative of $M$ in terms of its lower order derivatives. This can be achieved by means of the equations (\ref{T2}) and (\ref{T3}). In fact, if we consider the combination (\ref{T3})-$2$(\ref{T2}) we obtain
\[
\frac{M^{''}}{r}=\frac{2M^{'}}{r^2}-\frac{4M}{r^3}-\frac{2(M^{'}r-M)}{r^3}\left(\frac{M^{'}r-M}{r-2M}-1\right).
\]
Substitution of the above expression in (\ref{A}) gives (\ref{TOV}). Hence, the TOV equation is identically
satisfied for the case under consideration. In spite of this and the positive definite pressure, the model
displays one defect. As shown below the interior metric cannot be matched to the Schwarzschild metric 
\begin{equation} \label{xxx}
ds^2=\left(1-\frac{2\widehat{M}}{r}\right)dt^2-\left(1-\frac{2\widehat{M}}{r}\right)^{-1}dr^2-r^2(d\vartheta^2+\sin^2{\vartheta}d\varphi^2).
\end{equation}
with a total mass $\widehat{M}=M(R)$. In fact, continuity at the Schwarzschild radius $R$ requires that
\[
\frac{AR}{R-2\widehat{M}}=1+\frac{B-2M(R)}{R},\qquad h(R)=1-\frac{2\widehat{M}}{R}.
\]
The second condition implies that $B=0$ whereas the first equation implies that $A=(1-2\widehat{M}/R)^2$. However, there is a third condition \cite{Plebanski} that it has to be satisfied, namely
\[
A\sqrt{h(R)}\left.\frac{d}{dr}\left(\frac{1}{h}\right)\right|_{r=R}=\frac{2\widehat{M}}{R^2}\sqrt{1-\frac{2\widehat{M}}{R}}
\]
which gives $A=-(1-2\widehat{M}/R)^2$. Thus, we reached a contradiction. In the next section, we try to maintain the good features of the present model and extend it slightly such that the exterior Schwarzschild metric smoothly fits into the full solution.

\section{Anisotropic fluid solution}
We shall derive the complete solution of the gravitational field equations for a noncommutative geometry inspired anisotropic fluid described by a nonlocal equation of state. We consider again a spherically symmetric static distribution of matter $\rho_\theta$ but now the ansatz for the line element is
\begin{equation}\label{Schw}
ds^2=A^2(r)~dt^2-\frac{dr^2}{V(r)}-r^2(d\vartheta^2+\sin^2{\vartheta}~d\varphi^2).
\end{equation}
The energy-momentum tensor inside the matter distribution is supposed to be
\begin{equation}\label{aemt}
T^{\mu}_{\nu}=(\rho_\theta+P_\bot)u^{\mu}u_\nu-P_\bot\delta^\mu_\nu+(P_r-P_\bot)n^\mu n_\nu
\end{equation}
where $u^\mu$ is the velocity field of the fluid, $n^\mu=\sqrt{V}\delta^\mu_1$ is a unit spacelike vector in the radial direction, $P_r$ is the normal pressure, i.e. the pressure in the direction of $n^\mu$ and $P_\bot$ is the so-called 
tangential pressure, i.e. the pressure orthogonal to $n^\mu$. Since we consider a static matter distribution we have to require that $u^{i}=0$ for each $i=1,2,3$. Moreover, from $g_{\mu\nu}u^\mu u^\nu=1$ it follows that $u^0=1/A$. In this setting the energy-momentum tensor becomes
\[
T^{\mu}_{\nu}={\rm{diag}}(\rho_\theta,-P_r,-P_\bot,-P_\bot).
\] 
We shall assume that $P_r\neq P_\bot$ otherwise we would have again the case of an isotropic fluid. Notice that the quantity $\Delta=P_\bot-P_r$ called anisotropic factor is an indicator of the fluid anisotropy. Einstein field equations $G_{\mu\nu}=-8\pi T_{\mu\nu}$ become
\begin{eqnarray}
&&\frac{1-V}{r^2}-\frac{V^{'}}{r}=8\pi\rho_\theta,\label{unoa}\\
&&\frac{2VA^{'}}{Ar}+\frac{V-1}{r^2}=8\pi P_r,\label{duea}\\
&&\frac{1}{2rA}\left(2VA^{'}+AV^{'}+2rVA^{''}+rA^{'}V^{'}\right) =8\pi P_\bot.\label{trea}
\end{eqnarray}
On the other side the conservation equation $T^{\mu\nu}{}_{;\nu}=0$ with $\mu=r$ gives rise to the equation
\begin{equation}\label{quattroa}
P^{'}_r+\frac{A^{'}}{A}\left(P_r+\rho_\theta\right)=\frac{2}{r}(P_\bot-P_r).
\end{equation}
By means of equations (\ref{unoa}) and (\ref{duea}) it is not difficult to verify that equations (\ref{trea}) and (\ref{quattroa}) are equivalent, i.e. they are the same equation. In what follows we shall work with the system represented by (\ref{unoa}), (\ref{duea}) and (\ref{quattroa}). Although such a system is under-determined since we have three equations for the four unknown functions $A$, $V$, $P_r$ and $P_\bot$ it can be closed by assuming an equation of state for matter. In what follows we shall model the radial pressure by means of (\ref{NLES}) for $r\in[0,R]$ such that $P_r(R)=0$. According to the analysis performed in the previous section $R$ is the radius of the self-gravitating droplet. The nice feature is that the radial pressure is now positive inside the droplet. Outside the droplet we impose $P_r(r)=0$ for $r\geq R$. We use the freedom of the parameter $\theta$ to fix $R=r_s$. This allows a smooth
transition to the exterior Schwarzschild metric. In figure 1 we plot $\theta P_\bot$ versus $r^*$ to
explicitly demonstrate that the droplet has a finite extension.

\begin{figure}
\begin{center}
\scalebox{0.5}{\includegraphics{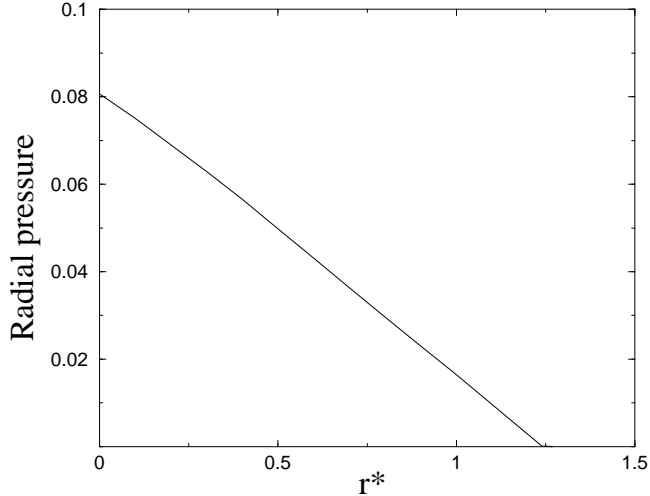}}
\caption{The radial pressure $\theta P_r$ versus $r^*=r/\sqrt{\theta}$}
\end{center}
\end{figure}

The function $V(r)$ can be obtained from (\ref{unoa}) and we have
\begin{equation}\label{V}
V(r)=1-\frac{2M(r)}{r}+\frac{C_1}{r}
\end{equation}
with $C_1$ integration constant and $M(r)$ given by (\ref{gamma}). Moreover, integrating (\ref{duea}) we find that
\[
A^2(r)=C_2 e^{\phi(r)},\quad \phi(r)=\int_0^r\psi(u)~du,\quad \psi(r)=\frac{1}{V(r)}\left(8\pi rP_r(r)+\frac{2M(r)-C_1}{r^2V(r)}\right).
\]
Thus, we have derived the line element
\begin{equation}\label{Schwan}
ds^2=C_2 e^{\phi}~dt^2-\frac{dr^2}{V(r)}-r^2(d\vartheta^2+\sin^2{\vartheta}~d\varphi^2)
\end{equation}
describing the manifold $\mathbb{R}\times[0,R]\times S^2$. Finally, the tangential pressure can be obtained from (\ref{quattroa}) and we have
\[
P_\bot(r)=P_r(r)+\frac{r}{2}P^{'}_r(r)+\frac{A^{'}(r)}{A(r)}(P_r(r)+\rho_\theta(r)).
\]
In figure 2 we plot this tangential pressure versus $r^*$ to show that it is also
positive definite. This part of the  pressure is non-zero
at the radius of the object (which is not a necessary
physical requirement), but has there a local minimum.

\begin{figure}
\begin{center}
\scalebox{0.5}{\includegraphics{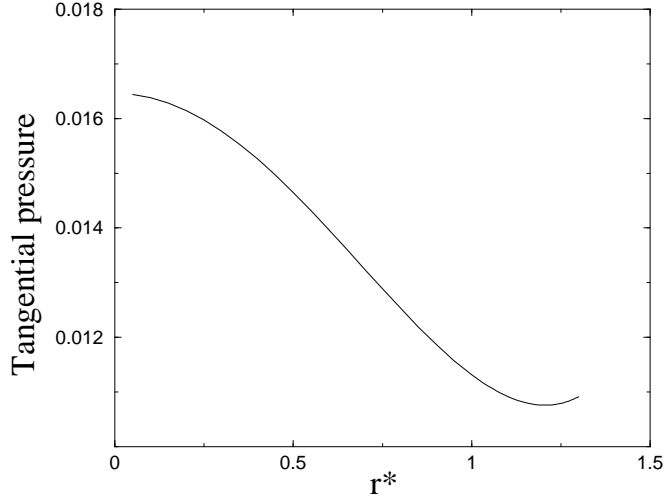}}
\caption{The tangential pressure $\theta P_\bot$ versus $r^*=r/\sqrt{\theta}$ for $\mu=2.2$}
\end{center}
\end{figure}

We match (\ref{Schwan}) with a Schwarzschild metric (\ref{xxx})
describing the outer region. In order to do that we have to require that $R$ the radius of the self-gravitating droplet equals the Schwarzschild radius. The matching of the line element inside the droplet with the Schwarzschild metric at the boundary $r=R$ is done by requiring the continuity of $A^2$ and $V$ at $r=R$. Thus, we find that
\[
C_1=0,\qquad C_2=\left(1-\frac{2M}{R}\right)e^{-\phi(R)}
\]
It is not difficult to verify that the line element we derived is flat at the center of the droplet.
The successful matching of the interior and exterior metric completes the program of
constructing a mini black hole solution inspired by non-commutative geometry.
We will briefly discuss the main features of the present model in the next section.

\section{Conclusions}
The model for a mini black hole presented in this paper is based on the
same basic physical principles as in \cite{Ansoldi}. 
As demonstrated in section 2 
the anisotropic energy-momentum tensor seems to be an unavoidable ingredient
in constructing self-gravitating droplets based on $\rho_{\theta}$.
The same anisotropic $T_{\mu \nu}$ has been used in \cite{Ansoldi}. 
However, we use a different equation of states (EOS), the so-called
non-linear EOS. The emerging self-gravitating droplet has then 
positive radial and tangential pressures in the interior,
a finite extension $R$ and the the interior solution can be
matched at $R$ with the exterior Schwarzschild geometry.
This seems to offer an alternative to the micro black hole
solution found in \cite{Ansoldi} where the radial and tangential pressures 
are negative and the droplet does not have a sharp finite radius.

\end{document}